\documentclass[amsmath,amssymb,amsfonts,aps,pre,preprint,superscriptaddress,bibnotes,showpacs,showkeys,longbibliography,dvipdfmx]{revtex4-1}
\usepackage{epsfig}
\usepackage[english]{babel}
\usepackage{color}
\usepackage{ulem}

\newcommand{\red}[1]{\textcolor{black}{#1}}

\begin{document}
\title{Three-phase equilibria in density-functional theory: interfacial tensions}
\author{Kenichiro Koga}
\affiliation{Research Institute for Interdisciplinary Science, Okayama University, Okayama 700-8530, Japan}
\affiliation{Department of  Chemistry, Faculty of Science, Okayama University, Okayama 700-8530, Japan}
\author{Joseph O. Indekeu}
\affiliation{Institute for  Theoretical Physics, KU Leuven, BE-3001 Leuven, Belgium}

\date{\today}

\begin{abstract}
A mean-field density-functional model for three-phase equilibria in fluids (or other soft condensed matter) with two spatially varying densities is analyzed analytically and numerically. The interfacial tension between any two out of three thermodynamically coexisting phases is found to be captured by a surprisingly simple analytic expression that has a geometric interpretation in the space of the two densities. The analytic expression is based on arguments involving symmetries and invariances. It is supported by numerical computations of high precision and it agrees with earlier conjectures obtained for special cases in the same model. An application is presented to three-phase equilibria in the vicinity of a tricritical point. Using the interfacial tension expression and employing the field variables compatible with tricritical point scaling, the expected mean-field critical exponent is derived for the vanishing of the critical interfacial tension as a function of the deviation of the noncritical interfacial tension from its limiting value, upon approach to a critical endpoint in the phase diagram. The analytic results are again confirmed by numerical computations of high precision. 
\end{abstract}

\maketitle

\section{Setting the stage}
  When three phases coexist in a $n$-component system, there remain
  $f=n-1$ degrees of freedom; i.e., there is then a $n-1$-dimensional
  manifold of triple points in a $n+1$-dimensional field space.
  The states in which two of the three coexisting phases become identical
  while in equilibrium with the third phase are 
  critical endpoints and those in which all the three become identical are
  tricritical points. 
  When $n=3$, or when $n>3$ with $n-3$ constraints, the manifold 
  of three-phase states is a surface, that of the critical endpoints is a curve, and that of the tricritical points is a point in the four-dimensional field space. There are two regimes of three-phase coexistence, one in which the three phases $\alpha$, $\beta$, and $\gamma$ meet at a common line of contact and the other in which one phase, say $\beta$, wets the $\alpha\gamma$ interface and there is no direct $\alpha\gamma$ contact. There may be a transition between
  these two regimes with changing thermodynamic state, called a wetting transition.

  There are pioneering experimental and theoretical studies on three-phase equilibria focusing on the associated critical phenomena and the structure and tension of interfaces~\cite{widom1973,Grif,LW,lang1976,widom1977,kerins1982, gama1983a, gama1983b}. Griffiths developed a phenomenological theory of the tricritical point and predicted the shape of the three-phase region in the density space~\cite{Grif}; Widom and co-workers experimentally determined the region of coexistence of three liquid phases verifying Griffiths' prediction~\cite{LW} and  
also gave an analytical expression for the variation of interfacial tensions as the three-phase region is traversed from one critical endpoint to the other, 
which results from a one-density van der Waals-Cahn-Hilliard theory~\cite{lang1976}; and Telo da Gama and Evans obtained the density profile and surface tension
near the critical endpoint of a binary mixture of Lennard-Jones fluids
based on a two-density density-functional theory (DFT)~\cite{gama1983a, gama1983b}. A two (or more) density theory is required in order to study
both the ``wet'' and ``nonwet'' regimes of three phase coexistence while any one-density theory cannot describe the nonwet regime.

Our aim in this contribution is to deepen and refine recent insights gathered from a mean-field density-functional theory (DFT) for  thermodynamic three-phase equilibria in condensed matter systems. In a recent paper Koga and Widom studied density-functional models of the structures and tensions of interfaces at three-phase equilibria close to a tricritical point \cite{KW}. In our present work we concentrate on a model, akin to model $T$ in Ref.~\cite{KW}, featuring a \red{local excess free-energy density} $F$ that is a product of three parabolic
\red{functions (``potential wells") of two \red{density variables} $\rho_1$ and $\rho_2$.} The \red{interfacial} free-energy density $\Psi$ in this model takes the form
\begin{equation}\label{Psi}
\Psi  = \frac{1}{2} \left (\frac{d\rho_1}{dz}\right )^2 +\frac{1}{2} \left(\frac{d\rho_2}{dz}\right )^2  + F(\rho_1,\rho_2)
\end{equation}
with
\begin{equation}\label{Fcomp}
F(\rho_1,\rho_2)= \prod_{\nu=\alpha,\beta,\gamma} V_{\nu}(\rho_1,\rho_2),
\end{equation}
where the potential wells take the form
\begin{equation}\label{V}
V_{\nu}(\rho_1,\rho_2)= (\rho_1 - \rho_{1}^{\nu} )^2 + (\rho_2 - \rho_{2}^{\nu})^2, \;\;\mbox{with}\;\; \nu = \alpha,\beta,\gamma,
\end{equation}
and the points $(\rho_{1}^{\alpha},\rho_{2}^{\alpha})$, $(\rho_{1}^{\beta},\rho_{2}^{\beta})$ and $(\rho_{1}^{\gamma},\rho_{2}^{\gamma})$ in the $(\rho_1,\rho_2)$-plane represent the densities of three coexisting bulk phases $\alpha$, $\beta$ and $\gamma$, respectively. 

Note that \red{$F$, along with $|\partial F/\partial \rho_1|$ and $|\partial F/\partial \rho_2|$, is zero at any of the densities of the three coexisting phases
  and otherwise positive} so as to describe a state of thermodynamic three-phase equilibrium. The \red{structure} of the interface between, say, phases $\alpha $ and $\gamma$, is then fully described by the two spatially varying densities $\rho_1(z)$ and $\rho_2(z)$, called density profiles, that depend on the \red{coordinate} $z$ perpendicular to the interface. \red{The equilibrium density profiles are those minimizing the interface free-energy functional
  $\int \Psi dz$ subject to the boundary conditions that the bulk phases infinitely far from the interface have the prescribed densities, e.g., $\rho_1(z)=\rho_1^\alpha$ and $\rho_2(z)=\rho_2^\alpha$ at $z= -\infty$ and $\rho_1(z)=\rho_1^\gamma$ and $\rho_2(z)=\rho_2^\gamma$ at $z=\infty$, and the interfacial tension
 of that interface is given by}
\begin{equation} \label{sigma}
\sigma_{\alpha\gamma} = \min_{\rho_1(z), \rho_2(z)} \int_{-\infty}^{\infty}\Psi (\rho_1,\rho_2)dz.
\end{equation}
The densities $\rho_1$ and $\rho_2$, the local and interfacial free energy densities $F$ and $\Psi$, the spatial coordinate $z$ are all taken to be dimensionless. To link the dimensionless quantities to experimental values, one may consider that energies are in units of $10^{-21}$~J and lengths in 1 nm. Then
the interfacial tensions will be in units of mN/m (=dyn cm$^{-1}$).\cite{kerins1982}

We recall that the equilibrium density profiles possess a useful first integral (or ``constant of the motion"),
\begin{equation} \label{com}
\frac{1}{2} \left (\frac{d\rho_1}{dz}\right )^2 +\frac{1}{2} \left(\frac{d\rho_2}{dz}\right )^2  = F(\rho_1,\rho_2).
\end{equation}

Our paper is organized as follows. In section 2 we  provide general results pertaining to interfacial tensions and in section 3 we turn our attention to the special case of systems in the vicinity of a tricritical point in the phase diagram. A brief conclusion and outlook on further applications close the paper. 

\section{Three-phase equilibria in a DFT with two densities.}
When three fluid phases $\alpha$, $\beta$, and $\gamma$ are in equilibrium,
the interfacial tensions $\sigma_{\alpha\gamma}$, $\sigma_{\alpha\beta}$, $\sigma_{\beta\gamma}$ for the $\alpha\gamma$, $\alpha\beta$, and $\beta\gamma$ interfaces
satisfy the triangle inequality \cite{RW}
\begin{equation}
\sigma_{\alpha\gamma} < \sigma_{\alpha\beta}  + \sigma_{\beta\gamma} 
\end{equation}
when the $\alpha\gamma$ interface is not wet by $\beta$ or the equality 
\begin{equation}
\sigma_{\alpha\gamma} = \sigma_{\alpha\beta}  + \sigma_{\beta\gamma} 
\end{equation} when the interface is wet by $\beta$ 
(with cyclic permutations and $\sigma_{ij} \equiv \sigma_{ji}$) \cite{review}.
In the former case the density profiles \red{for the non-wet $\alpha\gamma$ interface} directly connect the $\alpha$ and $\gamma$ phases in such a way that the interface trajectory in the $(\rho_1,\rho_2)$-plane connects the $\alpha$ and $\gamma$ phase points \red{without passing through the $\beta$ phase point}. In contrast, in the latter case the $\alpha\gamma$ trajectory passes through the $\beta$ phase point, which signifies that the $\beta$ phase intrudes in bulk between the $\alpha$ and $\gamma$ phases. The $\alpha\gamma$ interface then decomposes into two distinct interfaces separated by the bulk $\beta$ phase that wets it.

The interfacial tension of a non-wet interface can be computed from \eqref{sigma} using the equilibrium density profiles. \red{Alternatively, from \eqref{sigma} with \eqref{Psi}, \eqref{Fcomp}, and \eqref{com},
the tension, e.g., $\sigma_{\alpha\gamma}$, may be obtained from
\begin{equation}\label{sigma2}
\sigma_{\alpha\gamma}= 2 \int F(\rho_1,\rho_2) dz = 2 \int V_\alpha(\rho_1,\rho_2) V_\beta(\rho_1,\rho_2) V_\gamma(\rho_1,\rho_2) dz,
\end{equation}
with the same density profiles; note, however, that the integrals here are not extremal with respect to the density profiles.}
We now propose a method for analytically calculating the interfacial tension of a non-wet interface, e.g., $\sigma_{\alpha\gamma}$, within the two-density square-gradient density-functional theory of three-phase equilibria. The method, as illustrated in Fig.1, does not require knowledge of the actual density profiles. In our calculation the $\beta$ phase is considered to be a spectator phase with respect to the trajectory corresponding to the $\alpha\gamma$ interface (which, recall, does not pass through $\beta$). The potential well  $V_{\beta}$ associated with this spectator phase varies continuously within a range of positive values as the $\alpha\gamma$ interface trajectory runs from $\alpha$ to $\gamma$ in the $(\rho_1,\rho_2)$-plane. What we are proposing here 
is to find an effective constant potential $V^*_\beta$ that replaces $V_\beta$
in the original model and yet gives the same tension $\sigma_{\alpha\gamma}$
as the one that is found for the full $V_{\beta}$.


If the spectator well were a constant, the trajectory that solves the Euler-Lagrange equations would simply become the straight line connecting the $\alpha$ and $\gamma$ phase points in the $(\rho_1,\rho_2)$-plane. This is also conspicuous from symmetry considerations since the potential wells $V_{\alpha}$ and $V_{\gamma}$ are isotropic in the $(\rho_1,\rho_2)$-plane. The problem is then reduced to a single-density calculation, since $\rho_1$ and $\rho_2$ are linearly related with each other and their $z$-derivatives are proportional by a constant factor. Let  now $(\rho_1^*,\rho_2^*)$ be the density point in the $(\rho_1,\rho_2)$-plane such that $V_{\beta}^*\equiv V_{\beta}(\rho_1^*,\rho_2^*)$ is the effective constant value that the spectator well ought to take in order for the correct interfacial tension to be reproduced. With this $V_{\beta}^*$, from \eqref{sigma2}, we anticipate the following identity:
\begin{equation}\label{equiv}
\sigma_{\alpha\gamma}=2\int_{\rm actual}dz \;V_{\alpha}(\rho_1,\rho_2)V_{\beta}(\rho_1,\rho_2)V_{\gamma}(\rho_1,\rho_2) = 2V_{\beta}^*\int_{\rm straight}dz \;V_{\alpha}(\rho_1,\rho_2)V_{\gamma}(\rho_1,\rho_2),
\end{equation}
where the first integral involves the actual equilibrium density profiles, whereas the second integral only involves the simpler density profiles whose trajectory in the  $(\rho_1,\rho_2)$-plane lies on the straight line connecting $\alpha$ and $\gamma$. The challenge is to determine the point $(\rho_1^*,\rho_2^*)$ so that this equality holds for an arbitrary location of the $\beta$ phase point in the $(\rho_1,\rho_2)$-plane. 

The point $(\rho_1^*,\rho_2^*)$ that satisfies \eqref{equiv} can be calculated analytically for the special case when $\beta$ lies on the straight line through $\alpha$ and $\gamma$ in the $(\rho_1,\rho_2)$-plane, and outside the open line segment $(\alpha,\gamma)$. (If $\beta$ lay inside, the $\alpha\gamma$ interface would be necessarily wet by the $\beta$ phase, and so we excluded the possibility.). Let $\tau$ denote the coordinate along this line in the $(\rho_1,\rho_2)$-plane and let $\tau^{\alpha}\leq \tau^{\gamma}\leq \tau^{\beta}$ denote the bulk phase values in this density coordinate. With this setup the $\alpha\gamma$ interface is not wet by the $\beta$ phase while the $\alpha\beta$ interface is wet by the $\gamma$ phase. Note that the linear arrangement of three phase points in the $(\rho_1,\rho_2)$-plane necessarily realizes
  the wet regime of three phase equilibria as a one-density DFT model does. The integrals with respect to the spatial coordinate $z$ in \eqref{equiv} are transformed to those with respect to the density coordinate $\tau$ using a single-density analog of \eqref{com}. 
  The left-hand side of \eqref{equiv} is then analytically computed as

\begin{eqnarray}
\sigma_{\alpha\gamma}&=&\int_{\tau^{\alpha}}^{\tau^{\gamma}} d\tau\,\sqrt{2F_{1}(\tau)}\nonumber \\
&=& \sqrt{2}\int_{\tau^{\alpha}}^{\tau^{\gamma}} d\tau \,\left[\frac{(\tau^{\alpha}-\tau^{\gamma})^2}{4}-\left (\tau-\frac{\tau^{\alpha}+\tau^{\gamma}}{2}\right )^2\right] \left[\left (\tau^{\beta}- \frac{\tau^{\alpha}+\tau^{\gamma}}{2}\right) - \left(\tau - \frac{\tau^{\alpha}+\tau^{\gamma}}{2} \right )\right]\nonumber \\
&=& \frac{\sqrt{2}}{6}  (\tau^{\gamma}-\tau^{\alpha})^3\left (\tau^{\beta}- \frac{\tau^{\alpha}+\tau^{\gamma}}{2}\right),
\end{eqnarray}
with $F_{1}(\tau)$ similar to $F$ in \eqref{Fcomp} but for a single density.
In the long expression in these equations we have deliberately split the $\beta$-phase factor $\tau^{\beta} - \tau$ into its average value and its remainder, within the range of integration. This average value is precisely the effective constant we are looking for. Indeed, the remainder, being an odd function of $\tau-(\tau^{\alpha}+\tau^{\gamma})/2$ and being multiplied by the product of the $\alpha$ and $\gamma$-phase factors, which is even in $\tau-(\tau^{\alpha}+\tau^{\gamma})/2$, does not contribute to the integral.
{The right-hand side of \eqref{equiv} is $(\sqrt{2}/6)(\tau^\gamma-\tau^\alpha)^3 (\tau^\beta-\tau^*)$. Thus, we} conclude that the effective density is the midpoint between $\alpha$ and $\gamma$ in density space: 
\begin{equation}
\tau^*= (\tau^{\alpha} + \tau^{\gamma})/2
\end{equation}

We now conjecture that this is more generally true when $\alpha$, $\beta$ and $\gamma$ are not colinear in density space. To this end we first derive a general expression for the change of the interfacial tension under a variation of the bulk phase coordinates in the $(\rho_1,\rho_2)$-plane, while maintaining three-phase coexistence in bulk. Consider the following variations of the density profiles ($i=1,2$): $\rho_i (z)\rightarrow \rho_i (z)+ \delta \rho_i(z)$, caused by a variation of the bulk phase points ($\nu=\alpha,\beta,\gamma$): $\rho_{i}^{\nu}\rightarrow \rho_{i}^{\nu}+\delta \rho_{i}^{\nu}$. Then, to first order,
\begin{equation}
\delta \sigma = \int \; dz \sum _{i=1}^2 \left \{ \rho'_i (\delta \rho_i)'+ \frac{\partial F}{\partial \rho_i }\delta \rho_i+\sum _{\nu=\alpha}^{\gamma}  \frac{\partial F}{\partial \rho_i^{\nu} }\delta \rho_{i}^{\nu}\right \},
\end{equation}
where the prime denotes the $z$-derivative. \red{Since the equilibrium density profiles satisfy the coupled Euler-Lagrange equations, the contribution
  to the integral from the density variations $\delta\rho_i(z)$ is null, and so
\begin{equation}
\delta \sigma = \int \; dz \sum _{i=1}^2  \sum _{\nu=\alpha}^{\gamma}  \frac{\partial F}{\partial \rho_i^{\nu} }\delta \rho_{i}^{\nu}.
\end{equation}
This result is analogous to the Hellmann-Feynman theorem in molecular quantum
mechanics as remarked in Ref.~\cite{widom1979}.}
In order for $\sigma$ to be stationary with respect to shifts $ \delta \rho_{i}^{\nu} $ in the bulk phase points, we must meet the new requirement,
\begin{equation}\label{integralidentity}
\sum _{i=1}^2 \sum _{\nu=\alpha}^{\gamma} \delta \rho_{i}^{\nu} \int \; dz   \, \frac{\partial F}{\partial \rho_i^{\nu} } = 0.
\end{equation}
Consequently, the interfacial tension will be invariant only under special bulk phase point shifts that satisfy this constraint. 

In view of the special role played by the midpoint between $\alpha$ and $\gamma$ in density space, we suspect that $\sigma_{\alpha\gamma}$ may be invariant to shifts of the $\beta$ phase point that preserve its  distance to this midpoint (call it $\ell$; see Fig.~1). In other words, we investigate whether $\sigma_{\alpha\gamma}$ is invariant with respect to moving $\beta$ around, along a circle about this midpoint. Choosing the origin in the $(\rho_1,\rho_2)$-plane to be the midpoint of the straight line between $\alpha$ and $\gamma$, we designate the coordinates of the bulk phase points in the $(\rho_1,\rho_2)$-plane as $( \rho_{i}^{\alpha},0)$ for $\alpha$, $( -\rho_{i}^{\alpha},0)$ for $\gamma$ and $( \ell \cos\theta, \ell \sin\theta)$ for $\beta$.  We now perform an active rotation of the $\beta$ point about the midpoint by letting $\theta \rightarrow \theta+\delta \theta$. The condition \eqref{integralidentity} for this transformation takes the form
\begin{equation}\label{integralidentityOne}
\sin \theta  \int_{-\infty}^{\infty} \, dz   \, V_{\alpha}(\rho_1,\rho_2) V_{\gamma}(\rho_1,\rho_2) \rho_1 = \cos \theta \int_{-\infty}^{\infty} \, dz   \, V_{\alpha}(\rho_1,\rho_2) V_{\gamma} (\rho_1,\rho_2)\rho_2.
\end{equation}
\red{In polar coordinates $\rho_1 = r \cos\phi$, $\rho_2 = r \sin\phi$, this takes the form 
\begin{equation}\label{integralidentityOnepolar}
 \int _{-\infty}^{\infty}\, dz   \, V_{\alpha}(r,\phi) V_{\gamma}(r,\phi) r(z) \sin[\theta-\phi(z)]=0
\end{equation}
or, noting that $\sin(\theta-\phi)$ is the only factor in the integrand that changes its sign in the range of the integral, 
\begin{equation}\label{integralidentityOnepolarV2}
 \int _{\phi(z) < \theta}\, dz   \, V_{\alpha}(r,\phi) V_{\gamma}(r,\phi) r \sin(\theta-\phi)=- \int _{\phi(z) > \theta}\, dz   \, V_{\alpha}(r,\phi) V_{\gamma}(r,\phi) r \sin(\theta-\phi).
\end{equation}
Note that the integrand in \eqref{integralidentityOnepolar} is zero when $\phi=0, \theta, \pi$. 
}

A second integral identity can be derived by performing a passive rotation in which the $\beta$ point is fixed and the $\alpha$-$\gamma$ axis is rotated about its midpoint. In this case the coordinates of the bulk phase points in the $(\rho_1,\rho_2)$-plane are suitably chosen to be $( -R\cos\theta, R \sin\theta)$ for $\alpha$, $( R\cos\theta,-R \sin\theta)$ for $\gamma$ and $( \rho_1^{\beta},0)$ for $\beta$. For a variation $\theta \rightarrow \theta+\delta \theta$,  condition \eqref{integralidentity} now takes the form, 
\begin{equation}\label{integralidentityTwo}
\sin \theta  \int dz   [V_{\gamma}(\rho_1,\rho_2)-V_{\alpha}(\rho_1,\rho_2)]  V_{\beta}(\rho_1,\rho_2)\,\rho_1  = \cos \theta \int dz [V_{\alpha}(\rho_1,\rho_2)-V_{\gamma}(\rho_1,\rho_2)]  V_{\beta}(\rho_1,\rho_2)\,\rho_2. 
\end{equation}
In polar coordinates \red{$\rho_1 = r \cos(\phi-\theta)$, $\rho_2 = r \sin(\phi-\theta)$, where $0 \leq \phi \leq \pi$, 
\begin{equation}\label{integralidentityTwopolar}
  \int _{-\infty}^{\infty}\, dz   \, V_{\beta}(r,\phi) r^2(z) \sin[2\phi(z)]=0, 
\end{equation}
or 
\begin{equation}\label{integralidentityTwopolar2}
  \int _{0 < \phi(z) < \pi/2}\, dz   \, V_{\beta}(r,\phi) r^2 \sin(2\phi)=
  -\int _{\pi/2 < \phi(z) < \pi}\, dz   \, V_{\beta}(r,\phi) r^2 \sin(2\phi). 
\end{equation}
The integrand in \eqref{integralidentityTwopolar} is zero when $\phi=0, \pi/2, \pi$.}

Both the integral identities \eqref{integralidentityOne} and \eqref{integralidentityTwo} are obviously satisfied for two special configurations of the three-phase triangle in the $(\rho_1,\rho_2)$-plane. For $\theta =0$ (colinear configuration) the trajectories lie at $\rho_2=0$ so that both identities are trivially true. For $\theta = \pi/2$ (isosceles triangle) the antisymmetry in the density $\rho_1$  of the remaining integrand again ensures that the identities hold. For the general case we know of no analytic argument for proving the integral identities and we have recourse to numerical computations for the actual trajectories in arbitrary configurations. These confirm that the identities \red{\eqref{integralidentityOne}--\eqref{integralidentityTwopolar2}} hold numerically exactly for all $\theta$. 

We conclude that the interfacial tension (of a non-wet state) is invariant under a rotation of the $\beta$-phase point along a circle centered about the midpoint of the $\alpha$-$\gamma$ segment. Consequently, $\sigma$ indeed possesses the property \eqref{equiv} and the point  $(\rho_1^*,\rho_2^*)$ satisfies
\begin{equation}
\begin{cases}
\rho_1^*= (\rho_{1}^{\alpha} + \rho_{1}^{\gamma})/2\\
\rho_2^*= (\rho_{2}^{\alpha} + \rho_{2}^{\gamma})/2
\end{cases}
\end{equation}
This insight leads to a simple analytic expression for the interfacial tension, applicable to a three-phase triangle of general geometry, and for a non-wet interface, which reads
\begin{equation}\label{pcubeEll}
\sigma_{\alpha\gamma} = \frac{\sqrt{2}}{6} p^3 \ell,
\end{equation}
with $p$ the Euclidean distance from $\alpha$ to $\gamma$ and $\ell$ the Euclidean distance from $\beta$ to the midpoint of the $\alpha\gamma$ line, in the $(\rho_1,\rho_2)$-plane. The value of the interfacial tension calculated with this expression coincides with that computed by numerical integration to high precision using the full two-density model. Furthermore, for the special choices of three-phase equilibria (with isosceles three-phase triangles) for which analytic results for the interfacial tension were conjectured in \cite{KW}, our general expression is in accord with these conjectures. 

\section{Two-density DFT close to a tricritical point}
According to the theory of tricritical point phenomena two densities, $\rho_1$ and $\rho_2$, and two (dimensionless) field variables, say, $s$ and $t$ (e.g., linear combinations of temperature and pressure), are necessary and sufficient for describing the correct scaling properties of the thermodynamic quantities \cite{Grif,LW,RW}. For systems with three independent (chemical) components, the thermodynamic space is four-dimensional. 
\red{In a four dimensional field-variable space, three-phase states form a two-dimensional manifold (a surface), critical endpoints (CEPs), where two of the three phases become identical, form a one-dimensional manifold (a curve), and
the tricritical point is a single point.} In a neighborhood of the tricritical point, two lines of CEPs bound the surface of three-phase coexistence. The lines of CEPs merge tangentially, their distance vanishing algebraically with an exponent $3/2$ (in mean-field theory), as the tricritical point is approached. The same is true for the projections of these lines on the $(s,t)$-plane. In the vicinity of the tricritical point the two densities satisfy a scaling that selects out a principal density while the other becomes subsidiary. This implies a contraction of the projection of the three-phase triangle in the $(\rho_1,\rho_2)$-plane, upon approach of the tricritical point,  its sides becoming asymptotically aligned with the principal density. After defining suitable linear combinations of the densities, which we rename to be $\rho_1$ (principal) and $\rho_2$ (subsidiary), the scaling reads
\begin{equation}\label{rho2rho1}
\rho_2 = - \rho_1^2,
\end{equation}
which is intentionally coincident with the scaling adopted in model $T$ in \cite{KW}. In this representation the tricritical point is characterized by $\rho_1 = \rho_2 =0$. 

It is now possible to define linear combinations of the field variables, which we rename to be $s$ and $t$, so that the $t$-axis is parallel to the (common) asymptote of the lines of CEPs in the four-dimensional thermodynamic space and the approach of either one of the two CEPs at fixed $t$ is controlled by varying $s$. For concreteness, and without loss of generality, one may think of $t$ as being proportional to the temperature distance to the tricritical point. Varying  $s$ at constant $t$ can be thought of as varying the pressure  in order to interpolate between the two critical endpoints at constant temperature. In this representation the principal densities associated with the three coexisting phases, for given $s$ and $t$ in the three-phase coexistence range, are the zeroes of the third-degree polynomial
\begin{equation}\label{phi}
\phi (\rho_1) = \rho_1^3-3t\rho_1+2s
\end{equation}
The tricritical point is at $t=0$ and the three-phase coexistence range for $s$ is 
 \begin{equation}
-t^{3/2} \leq s \leq t^{3/2},
\end{equation}
the equalities being achieved in the respective CEPs.
The zeroes of \eqref{phi}, $a(s,t)\equiv \rho_1^{\alpha}(s,t)$, $b(s,t)\equiv \rho_1^{\beta}(s,t)$ and $c(s,t)\equiv \rho_1^{\gamma}(s,t)$, satisfy
\begin{equation}
\begin{cases}
-3t = ab+bc+ca\\
-2s = abc\\
  \;\;\;\;0 = a+b+c
\end{cases}
\end{equation}
The CEP densities are solutions of $d\phi/d\rho_1 =0$, which leads to
\begin{equation}
\begin{cases}
\rho_1^{\alpha}=\rho_1^{\beta}=-t^{1/2}\;\;\mbox{and} \;\; \rho_1^{\gamma}=2t^{1/2}\;\;\mbox{at} \;\;  \alpha\beta \;\; \mbox{criticality}\\
\rho_1^{\beta}=\rho_1^{\gamma}=t^{1/2}\;\;\mbox{and} \;\; \rho_1^{\alpha}=-2t^{1/2}\;\;\mbox{at} \;\;  \beta\gamma \;\; \mbox{criticality}
\end{cases}
\end{equation}

\red{
From \eqref{rho2rho1} with $\rho_1=a$, $b$, and $c$ at the three
coexisting phases $\alpha$, $\beta$, and $\gamma$,
the vertices of a triangle in the ($\rho_1$, $\rho_2$) plane in Fig. 1 are
located at $(a, -a^2)$, $(b, -b^2)$, and $(c, -c^2)$. Evaluating $p$ and $\ell$
from the coordinates and substituting them in \eqref{pcubeEll}, one has
an analytical expression for the interfacial tension near the tricritical point: e.g., 
\begin{equation}\label{sigma_ag}
  \sigma_{\alpha\gamma} = \frac{\sqrt{2}}{6} |c-a|^3\left[1+(a+c)^2\right]^{3/2}
  \left[ \left(\frac{a+c}{2}-b\right)^2 + \left(b^2-\frac{a^2+c^2}{2}\right)^2\right]^{1/2}
\end{equation}
for the $\alpha\gamma$ interface. Similarly $\sigma_{\alpha\beta}$ and $\sigma_{\beta\gamma}$ are given by cyclic permutations of $a$, $b$, and $c$ in the above expression. We now apply the general result \eqref{pcubeEll} or the above formula \eqref{sigma_ag} to several cases of interest.
}
\subsection{Interfacial tension at a critical endpoint}
At $\beta\gamma$ criticality ($s = t^{3/2}$) we calculate the tension $\sigma_{\alpha,\beta\gamma}$ of the interface between the non-critical phase $\alpha$ and the critical phase $\beta\gamma$ that results when $\beta$ becomes identical to $\gamma$. Likewise, $\sigma_{\alpha\beta, \gamma}$ can be obtained at $\alpha\beta$ criticality ($s = -t^{3/2}$). A simple exact calculation is possible in this case because there are only two coexisting phases and the trajectory in the $(\rho_1,\rho_2)$-plane is the straight line  $\rho_2 = t^{1/2} (\rho_1 - t^{1/2}) -t$. The result is
\begin{equation}\label{sigmaCEP}
\sigma_{\alpha,\beta\gamma}= \sigma_{\alpha\beta,\gamma}= \frac{27\sqrt{2}}{4} t^2(1+t)^2. 
\end{equation}
We remark that the same result is obtained by applying \eqref{pcubeEll} to the three-phase triangle approaching the CEP (where one of the edges vanishes). Furthermore, this result is a refinement of the conjecture proposed in equation 22 in \cite{KW} in the context of  model $T$. The explicit field dependence of the interfacial tension, obtained in \eqref{sigmaCEP}, allows one to read off the tricritical exponent $\mu_t= 2$ which characterizes the vanishing of the interfacial tension upon approach of the tricritical point ($t \rightarrow 0$).
\subsection{Critical interfacial tension close to a critical endpoint}
Close to $\beta\gamma$ criticality ($s \lesssim t^{3/2}$) we calculate the near-critical tension $\sigma_{\beta\gamma}$ of the diffuse $\beta\gamma$ interface that arises when $\beta$ is almost identical to $\gamma$. It is convenient to scale the densities with $t^{1/2}$ and to consider a small deviation from the CEP by setting $s = t^{3/2}(1-\epsilon)$, with $\epsilon$ small compared to unity. We then obtain the expansion
\begin{equation}
\begin{cases}\label{bulkEps}
t^{-1/2}a = -2 + \frac{2}{9} \epsilon + {\cal O}(\epsilon^2)\\
t^{-1/2}b = 1 - \sqrt{\frac{2}{3}} \;\epsilon^{1/2} + {\cal O}(\epsilon)\\
t^{-1/2}c = 1 + \sqrt{\frac{2}{3}} \;\epsilon^{1/2} + {\cal O}(\epsilon)
\end{cases}
\end{equation}
For $\epsilon \rightarrow 0$, approaching the CEP by varying $s$ at constant $t$, the interface trajectory from $\beta$ to $\gamma$ in the $ (\rho_1,\rho_2)$-plane becomes a straight line of vanishing length, the slope of which converges to $d\rho_2/d\rho_1 = -2t^{1/2}$. This allows one to perform an exact calculation in this limit. An expansion in $\epsilon$ leads to the analytic expression
\begin{equation}\label{sigmanearCEP}
\sigma_{\beta\gamma} =  \frac{16\sqrt{3}}{9} t^2(1+t)^{1/2}(1+4t)^{3/2}\epsilon^{3/2} + {\cal O}(\epsilon^2).
\end{equation}
\red{The same result is obtained from the analog of \eqref{sigma_ag}
for the $\beta\gamma$ interface.}
Since $\epsilon$ is linear in the deviation of the field $s$ from its value at the CEP, we conclude that the mean-field critical exponent for the interfacial tension is $\mu_c = 3/2$ as expected in ordinary mean-field theory for critical phenomena. Note that we may also choose a path towards the tricritical point ($t\rightarrow 0$) at constant $\epsilon$, for instance at $\epsilon =1$ (i.e., $s=0$), in which case we retrieve $\mu_t= 2$.
If, in \eqref{sigmanearCEP}, we choose to express the interfacial tension in terms of the difference in one of the densities of the near-critical phases, we obtain, using \eqref{bulkEps},
\begin{equation}\label{sigmanearCEPd}
\sigma_{\beta\gamma} \propto (\rho_1^{\gamma}-\rho_1^{\beta} )^3,
\end{equation}
featuring the familiar third power that is expected in mean-field theory for the approach to a critical point. Similar results evidently hold for the interfacial tension $\sigma_{\alpha\beta}$ near the other CEP. The results of this subsection are all in accord with those obtained for model $T$ in \cite{KW}.

\subsection{General interfacial tension for a non-wet interface}


Within the present DFT model, the interfacial tension for a non-wet interface
is given by \eqref{sigma_ag} or its analog. 
As we already remarked in the foregoing section, this analytic form is in agreement with high-precision numerical computation using the full model. We now proceed to use it to make an analytic prediction for the singularity displayed by the {\em non-critical} interfacial tension $\sigma_{\alpha\beta}$ as it approaches the value $\sigma_{\alpha,\beta\gamma}$ near the CEP where $\beta$ and $\gamma$ become identical to a common critical phase.
\red{From the analog of \eqref{sigma_ag} for the $\alpha\beta$ interface,} 
again setting $s = t^{3/2}(1-\epsilon)$, with $\epsilon$ small compared to unity, we obtain 
\begin{equation}\label{CEPdev}
\sigma_{\alpha\beta} -  \sigma_{\alpha,\beta\gamma} \propto \epsilon +  {\cal O}(\epsilon^{3/2}).
\end{equation}
The exponent of $\epsilon$, and therefore of the distance in the field $s$ from the CEP, in the leading term takes the value unity. This is in accord with the theoretical prediction that the non-critical interfacial tension approaches its value at the CEP in a manner that is linear in the fields. Scaling theory predicts an additional singular contribution to the non-critical interfacial tension as a function of the field variable (in our case $\epsilon$), with an exponent $\mu_c$, being the critical exponent of the interfacial tension at criticality \cite{widom1985,FU,MEF}. In mean-field theory that exponent takes the value $3/2$ and the experimental value for $\mu_c$ is around 1.3 for three-dimensional fluids \cite{SW}. The theoretical value (beyond mean-field theory) in three dimensions is near 1.26. In any case, the singular contribution is a correction of higher order than the leading term because $\mu_c >1$.

Note that \eqref{CEPdev} implies that, in mean-field theory, the leading deviation is quadratic in the density difference of the critical phases. We obtain 
\begin{equation}\label{CEPdens}
\sigma_{\alpha\beta} -  \sigma_{\alpha,\beta\gamma} \propto (\rho_1^{\gamma}-\rho_1^{\beta} )^2 + {\cal O}((\rho_1^{\gamma}-\rho_1^{\beta} )^3)
\end{equation}

In view of  \eqref{sigmanearCEPd} this implies the following generic relationship between the interfacial tensions of the non-wet interfaces
\begin{equation}\label{CEPrel}
\sigma_{\beta\gamma} \propto (\sigma_{\alpha\beta} -  \sigma_{\alpha,\beta\gamma} )^{3/2},
\end{equation}
which features the exponent $3/2$ expected in mean-field theory, while the experimental value is about 1.3. \red{Figure 2 illustrates variation of the $\alpha\beta$ and $\beta\gamma$ interfacial tensions at a fixed $t$ and the relationship \eqref{CEPrel} between the vanishing tension $\sigma_{\beta\gamma}$ and the vanishing difference $\sigma_{\alpha\beta} - \sigma_{\alpha,\beta\gamma}$ on approach to the $\beta\gamma$ critical endpoint, as obtained by \eqref{pcubeEll} or analogs of \eqref{sigma_ag}}. The reason that an earlier model, model T in \cite{KW}, gave the value 3, instead of 3/2, for this exponent is seen in our calculations to be due to the particular way the CEPs are approached in model T, treating $\rho_1^\alpha$, $\rho_1^\beta$, $\rho_1^\gamma$ as independent variables and letting $\rho_1^\beta$ go to either fixed $\rho_1^\alpha$ or fixed $\rho_1^\gamma$.
  In contrast, in our model the bulk densities are constrained by the physical requirement that they be solutions of \eqref{phi} for given field values $s$ and $t$. In closing this subsection we note that in exceptional circumstances within our model an exponent different from the generic value $3/2$ may be obtained (notably the value $3/4$). We postpone a discussion of this until future work.

\section{Conclusion and outlook}
We have studied by means of analytical calculations, and supported by numerical computation, the interfacial tension in a density-functional theory for three-phase equilibria employing two spatially varying density profiles. Within a model DFT that features a free-energy density that is a product of three isotropic potential wells in density space, we have derived a simple analytic expression for the interfacial tension between two phases (that are not wet by the third phase). This expression, which has been obtained through analytic arguments and proven to be numerically exact by direct computation, possesses a simple geometrical interpretation in the two-dimensional density space. 

Applying the DFT to the vicinity of a tricritical point in thermodynamic space, we have expressed the bulk phase densities and the interfacial tension in terms of the field variables that satisfy the correct scaling requirements near tricriticality. In this way we have obtained analytically, and verified numerically, that the generically expected mean-field critical exponent is obtained for the critical interfacial tension near a critical endpoint as a function of the difference between the noncritical interfacial tension and its critical-endpoint value. 

We consider to apply the handy analytical results obtained in this work to map out efficiently and in detail the global wetting phase diagram for this DFT close to, and also away from, the tricritical point. The wetting phenomena within this DFT have to some extent been studied previously \cite{KW} and we hope to refine and complete this investigation in future work.

\section*{Acknowledgements} 
We thank Professor Benjamin Widom for helpful comments. 
JOI thanks Okayama University for hospitality and the Japan Society for the Promotion of Science for a JSPS Invitational Fellowship for Research in Japan with ID No. S18131. KK acknowledges support by JSPS KAKENHI Grant No. 26287099
and No. 18KK0151. 

{}

\newpage
\begin{figure}[h!]
\centering
\includegraphics[width=0.9\textwidth]{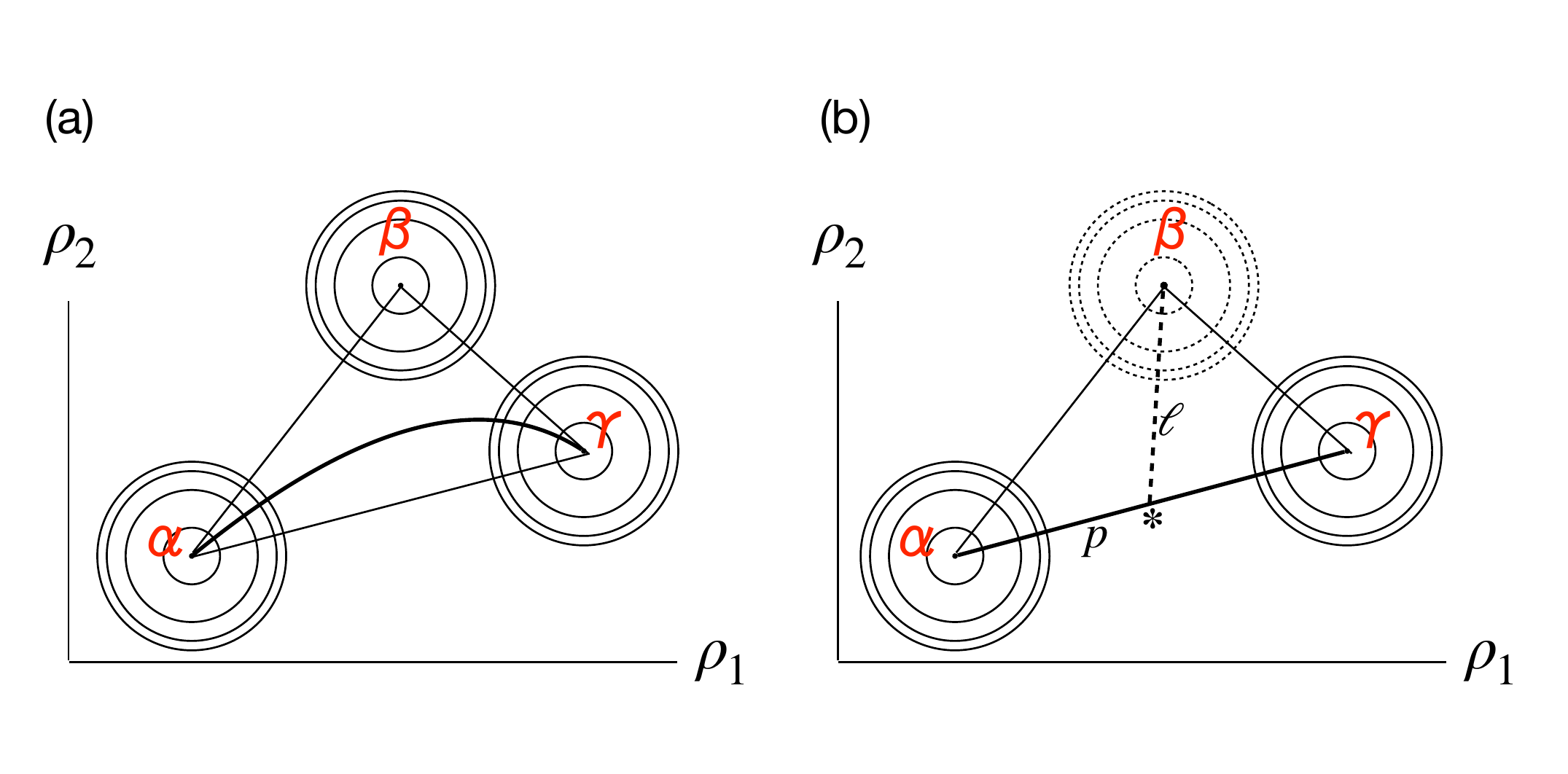}
\caption{Equivalence of two ways of obtaining the interfacial tension $\sigma_{\alpha\gamma}$ of a non-wet interface in a three-phase equilibrium. (a) Three-phase triangle in the plane of two densities $\rho_1$ and $\rho_2$. The bulk phase points are indicated by $\alpha$, $\beta$ and $\gamma$. Near those points contours of constant free-energy density $F(\rho_1,\rho_2)$ are circular \red{as depicted by circles}. The trajectory corresponding to the $\alpha\gamma$-interface structure in the full DFT model is \red{schematically shown by the thick solid curve connecting the $\alpha$ and $\gamma$ points.}  The interfacial tension $\sigma_{\alpha\gamma}$ is computed based on this trajectory. (b) A simplified model in which the potential well $V_{\beta}(\rho_1, \rho_2)$ associated with the spectator phase $\beta$ in $F$ has been suppressed (as indicated by dotted circles) and replaced by a constant factor in $F$. Under these circumstances the trajectory corresponding to the $\alpha\gamma$-interface becomes straight (thick solid line). Adjusting the constant factor to the value that $V_{\beta}(\rho_1, \rho_2)$ takes in the midpoint of the trajectory, indicated by the star, reproduces the interfacial tension computed in (a). The calculation in (b) leads to an analytic expression for $\sigma_{\alpha\gamma}$ which is proportional to the third power of the length $p$ of the edge $\alpha\gamma$ in the triangle $\alpha\beta\gamma$ and the length $\ell$ of the median connecting $\beta$ to the principal edge.}
\label{1}
\end{figure}

\begin{figure}[h!]
\centering
\includegraphics[width=0.9\textwidth]{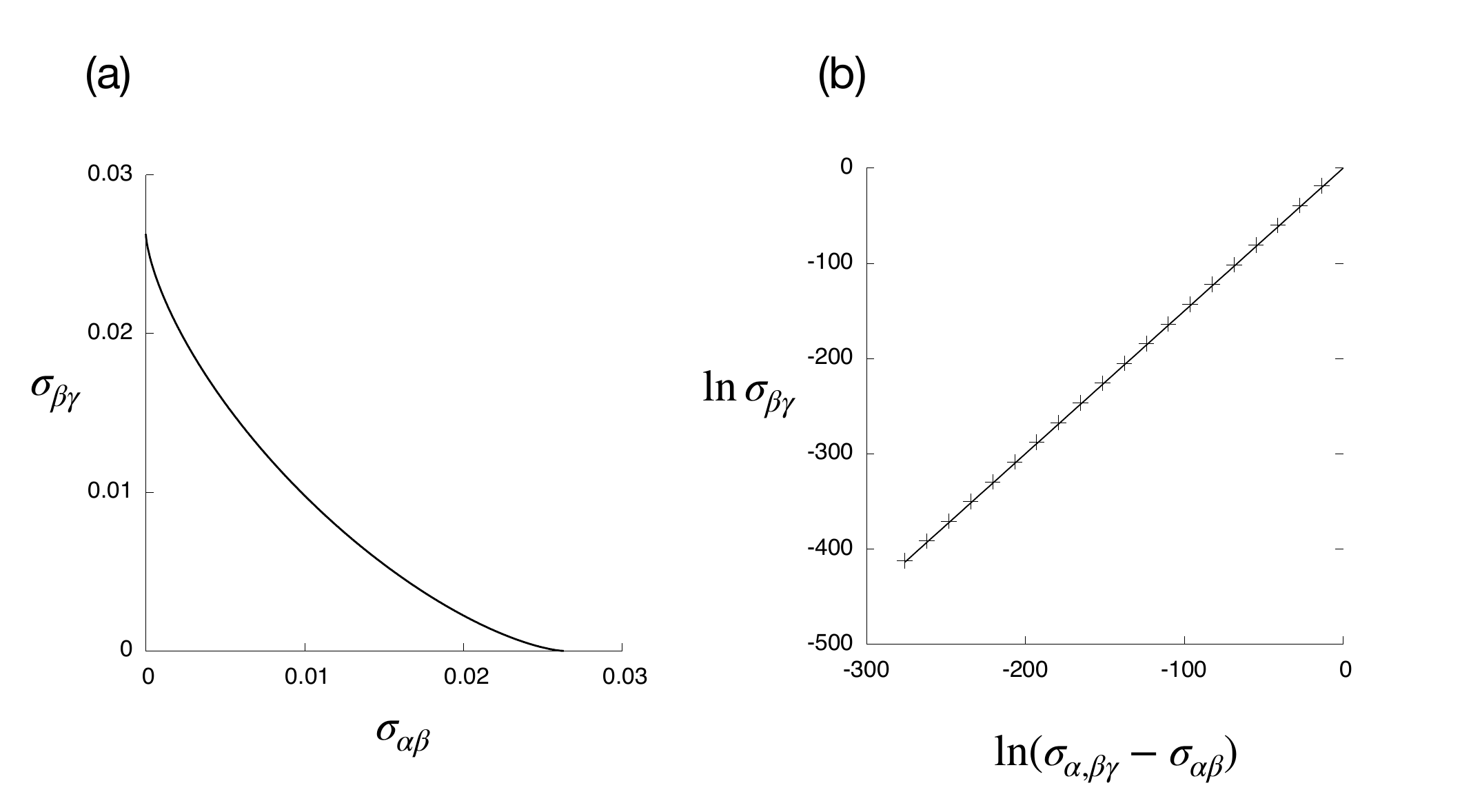}
\caption{
  \red{Variations of the $\alpha\beta$ and $\beta\gamma$ interfacial tensions at fixed $t$. (a)  $\sigma_{\beta\gamma}$ versus $\sigma_{\alpha\beta}$.
  (b) Log-log plot of $\sigma_{\beta\gamma}$ versus $(\sigma_{\alpha,\beta\gamma}-\sigma_{\alpha\beta})$. The curve in (a) and points in (b) are numerical data obtained from the analogs of \eqref{sigma_ag}. The field variable $t$ is fixed at 0.05 while $s$ varies between $-t^{3/2}$ and $t^{3/2}$. The slope of the line in (b) is 3/2, confirming the 3/2-power tangency at the critical endpoints.} }
\label{2}
\end{figure}

\end{document}